# Large Scale Structures as Gradient Lines:
# The Case of Trkal Flows

A. Libin

**Abstract**

**A specific asymptotic expansion at large Reynolds numbers *(R)* for the long wavelength perturbations of non stationary anisotropic helical solutions of the forceless Navier-Stokes equations (Trkal solutions) is efficiently constructed of the Beltrami type terms through multi scaling analysis. The asymptotic procedure is proved to be valid for one specific value of the scaling parameter, namely $R^{1/2}$. As a result, large scale structures arise as gradient lines of the energy density determined by the initial conditions for two anisotropic Beltrami flows of the same helicity.** The same initial conditions determine the boundaries of the vortex- velocity tubes, containing both streamlines and vortex lines.

**Introduction**

The idea that the coherent structures in the developed turbulence are in fact the domains of the Beltrami-like flows appeared in the pioneering works of E. Levich and co-authors, ([4-6],[25]), and in seminal works of H.K. Moffat [15-17]. The emergence in the developed turbulence of domains with helical, Beltrami-like flows, distinguished by the alignment of the velocity and the vorticity vectors, is presently firmly established in geophysical observations, as well as in direct numerical simulations of the Navier Stokes equations ([2],[3],[10-14],[18],[19],[20]).

In view of all this, in the present work the existence of the Beltrami like domains in turbulent flows is preordained.. According to the concept of E. Levich [6], the Kolmogorov spectrum [K41] is likely to be generated by a particular class of sub-domains with Beltrami flows inside them that are embedded in the midst of a turbulent flow domain. At the core of this concept lies the idea of an ensemble of emerging and dying helical fluctuations. A remarkable property of the helical fluctuations is that they either occupy very small sub-domains, actually sub-domains having fractal dimensions equal or less than 2.5, or have very little life time span, so their 4-D phase volume always remains small. During the life time of Beltrami flows fluctuations, the nonlinear term in the Navier-Stokes equations vanishes and this results in a global reduction of the nonlinear interaction in the whole fluid volume. But the helical fluctuations emerging and dying are still not the coherent structures that have seemingly an unlimited life time span, which are observed experimentally.

To understand the nature of CS we apply the ideas and mathematical tools invented by G.I. Sivashinsky [23]. In this work he considered formation of large scale structures as a manifestation of long wavelength instability of space periodic solutions of the Navier Stokes equations. In particular, this idea was tackled in G. Sivashinsky and co-authors ([8-9],[21],[22]). In these works the authors studied instability of the *linearised* Navier-Stokes equations, rather than the equations themselves. Sometimes they implied in these studies the Galerkin approximations. In another work by A. Libin, G.I. Sivashinsky and E. Levich [7] the authors studied long wavelength linear instability of the non-stationary Trkal [24] solution for the force free Navier-Stokes equations and the time independent solution for the forced Navier-Stokes equations at large values of the Reynolds number $R$. It was concluded that while the *linearised* forced Navier-Stokes equations are unstable to the perturbations of the characteristic wavelength proportional to $R$, the non stationary solution of the *linearised* forceless Navier-Stokes equation might be unstable to the perturbations of some intermediate large wavelength $L$, smaller than $R$. This intermediate large scale $L$, remains undetermined.

In the present paper the nonlinear asymptotic scaling analysis for the long wavelength perturbations of the Trkal solution of the force free Navier-Stokes equations is performed and their specific asymptotic solution, *constructed of Beltrami-like terms* is effectively built, while the whole procedure proved to be valid for one and only specific scale (intermediate wavelength), which equals $R^{1/2}$. This brings the explicit large scale streamline equations. Quasi-stationary solutions of these equations could be considered as large scale structures. The crucial point of the analysis proves to be the coupling of two large scale amplitude modulated Beltrami flows of the same vorticity.

There is a striking similarity between the necessity of the $R^{1/2}$ scaling due to asymptotic expansion in Beltrami-like terms in the present work and the reduction by $R^{1/2}$ of the coupling constant in front of the nonlinear term in asymptotically convergent renormalized perturbation expansion developed in E. Levich [6], based on the concept long advanced by K.H. Moffatt and E. Levich in the papers cited above, claiming that developed turbulent flows are dominated by the helical Beltrami-like fluctuations. As in the present theory the reduced value of the coupling constant in [6] is unambiguous and only one for the asymptotic convergence of the theory. Incidentally the renormalized coupling parameter is directly connected with the prevalence of the Kolmogorov law in a part of the energy spectrum in the momentum space of wave numbers [6].

## 1 General Framework of Scaling

We consider the non-forced Navier-Stokes equation in rotational and non dimensional form for the incompressible and viscous fluids:

$$\begin{cases} \dfrac{\partial}{\partial t}(\operatorname{rot}\vec{u}) + \operatorname{rot}[\operatorname{rot}\vec{u} \times \vec{u}] = \dfrac{1}{R}\Delta(\operatorname{rot}\vec{u}) \\ \operatorname{div}\vec{u} = 0 \end{cases} \qquad (1)$$

These equations have the so-called Trkal solutions [24]:

$$\vec{u}_0(\vec{r},t) = A e^{\frac{-\lambda^2}{R}t} \cdot \vec{h}(\vec{r}),$$

where $\vec{h}(\vec{r})$ is a helical flow:

$$\operatorname{rot}\vec{h}(\vec{r}) = \lambda \vec{h}(r)$$

The existence of a continuous helical flow means some form of periodicity in space.

We consider a specific anisotropic solution with $\lambda = 1$.

$$\vec{h}_0(z) = \begin{pmatrix} \sin z \\ \cos z \\ 0 \end{pmatrix} \quad ; \quad \vec{u}_0(z,t) = Ae^{\frac{-t}{R}} \cdot \vec{h}_0(z)$$

One can easily check that

$$\text{rot}\,\vec{h}_0(z) = \vec{h}_0(z)$$

Our aim is the investigation of the long-wavelength instability of this solution for equation (1) at large values of the Reynolds numbers

$$R \gg 1$$

Consider $\vec{u} = \vec{V} + \vec{u}_0$. Then, the equation for $V$ is

$$\begin{cases} \dfrac{\partial}{\partial t}(\text{rot}\,\vec{V}) + Ae^{-t/R}\text{rot}[(\text{rot}\,\vec{V} - \vec{V}) \times \vec{h}_0(z)] + \text{rot}[\text{rot}\,\vec{V} \times \vec{V}] = \dfrac{1}{R}\Delta \text{rot}\,\vec{V} \\ \text{div}\,\vec{V} = 0, R \gg 1 \end{cases} \quad (2)$$

The idea is the existence of slowly evolving large scale amplitude modulations of $\vec{V}$.

Following G. Sivashinsky [23], in order to discern these modulations, we perform a formal scaling procedure:

$$\vec{V}(x,y,z,t) \longrightarrow \vec{V}(x,y,z,\xi,\eta,\theta,\tau)$$

$$\frac{\partial}{\partial x} \longrightarrow \frac{\partial}{\partial x} + \varepsilon \frac{\partial}{\partial \xi} \quad (\xi \sim \varepsilon x)$$

$$\frac{\partial}{\partial y} \longrightarrow \frac{\partial}{\partial y} + \varepsilon \frac{\partial}{\partial \eta} \quad (\eta \sim \varepsilon y)$$

$$\frac{\partial}{\partial z} \longrightarrow \frac{\partial}{\partial z} + \varepsilon \frac{\partial}{\partial \theta} \quad (\xi \sim \varepsilon z)$$

$$\frac{\partial}{\partial t} \longrightarrow \varepsilon \frac{\partial}{\partial \tau} \quad (\tau = \varepsilon t) \quad (3)$$

where $\varepsilon$ is some small parameter,

$$\text{rot}_{xyz} \longrightarrow \text{rot}_{xyz} + \varepsilon \text{rot}_{\xi\eta\theta}$$

$$\text{div}_{xyz} \longrightarrow \text{div}_{xyz} + \varepsilon \text{div}_{\xi\eta\theta}$$

In fact, we shall seek for solutions of the scaled equations that are not dependent in $x$, $y$ and $\theta$.

Thus, (2) is transformed into:

$$Ae^{-\varepsilon\tau}\text{rot}_z[(\text{rot}_z\vec{V}-\vec{V})\times\vec{h}_0]+\text{rot}_z[\text{rot}_z\vec{V}\times\vec{V}]+$$
$$+\varepsilon\left\{Ae^{-\varepsilon\tau}\left\{\text{rot}_z[\text{rot}_{\xi\eta}\vec{V}\times\vec{h}_0]+\text{rot}_{\xi\eta}[(\text{rot}_z\vec{V}-\vec{V})\times\vec{h}_0]\right\}+\right.$$
$$\left.+\text{rot}_{\xi\eta}[\text{rot}_z\vec{V}\times\vec{V}]+\text{rot}_z[\text{rot}_{\xi\eta}\vec{V}\times\vec{V}]+\frac{\partial}{\partial\tau}(\text{rot}_z\vec{V})-\frac{1}{R}\text{rot}_{\xi\eta}\Delta_z\vec{V}\right\}+$$
$$+\varepsilon^2\left\{\text{rot}_{\xi\eta}[\text{rot}_{\xi\eta}\vec{V}\times\vec{V}]+Ae^{-\varepsilon\tau}\text{rot}_{\xi\eta}[\text{rot}_{\xi\eta}\vec{V}\times\vec{h}_0]+\frac{\partial}{\partial\tau}(\text{rot}_{\xi\eta}\vec{V})+\right.$$
$$\left.+\frac{1}{R}\text{rot}_z\Delta_{\xi\eta}\vec{V}\right\}-\varepsilon^3\cdot\frac{1}{R}\Delta_{\xi\eta}\text{rot}_{\xi\eta}\vec{V}-\frac{1}{R}\cdot\Delta_z\text{rot}_z\vec{V}=0 \quad (4)$$
$$\text{div}_z\vec{V}+\varepsilon\text{div}_{\xi\eta}\vec{V}=0$$

The term $\Delta_z\text{rot}_z V$ should be one of the terms in (4), thus, for some integer $k$, we have

$$\varepsilon^k=R^{-1} \quad \text{or} \quad \varepsilon=R^{-\frac{1}{k}}$$

We shall seek the solution of (4) as asymptotic expansion in powers of $\varepsilon$:

$$\vec{V}\sim\vec{V}_0+\varepsilon\vec{V}_1+\varepsilon^2\vec{V}_2+\cdots \quad (5)$$

Set

$$\vec{h}_1=\begin{pmatrix}\cos z\\-\sin z\\0\end{pmatrix}$$

Obviously

$$\text{rot}\,\vec{h}_1=\vec{h}_1$$

There are no other linear independent $2\pi$-periodic in $z$ flows of unit helicity except $\vec{h}_0$ and $\vec{h}_1$.

We shall seek the terms of the asymptotic expansion (5) as:

$$\vec{V}_n=\gamma_0^{(n)}(\xi,\eta,\tau)\vec{h}_0(z)+\gamma_1^{(n)}(\xi,\eta,\tau)\vec{h}_1(z)+$$
$$+\vec{\vartheta}_n(\xi,\eta,z,\tau)+\vec{\delta}_n(\xi,\eta,\tau)=\vec{w}_n+\vec{\vartheta}_n+\vec{\delta}_n \quad (6)$$

where

$$\vec{\vartheta}_n=\begin{pmatrix}0\\0\\B_n\cos z+C_n\sin z\end{pmatrix} \quad ; \quad \vec{\delta}=\begin{pmatrix}0\\0\\\delta_n(\xi,\eta,\tau)\end{pmatrix}$$

Clearly

$$\text{rot}_z \vec{g}_n = 0$$

$$\text{rot}_z \vec{V}_n = \text{rot}_z[\gamma_0^{(n)}\vec{h}_0(z) + \gamma_1^{(n)}\vec{h}_1(z)] = \text{rot}_z \vec{w}_n = \vec{w}_n$$

**We shall effectively construct the expansion (5) in form (6).**

Since we are seeking for $2\pi$ - periodic in $z$ solutions of (4), we may integrate (4) over the period:

$$\text{rot}_{\xi\eta}\int_0^{2\pi}\left\{Ae^{-\varepsilon\tau}[(\text{rot}_z\vec{V}-\vec{V})\times\vec{h}_0] + [\text{rot}_z\vec{V}\times\vec{V}]\right\}dz$$
$$+\varepsilon\text{rot}_{\xi\eta}\int_0^{2\pi}\left\{[\text{rot}_{\xi\eta}\vec{V}\times(\vec{V}+Ae^{-\varepsilon\tau}\vec{h}_0)] + \frac{\partial\vec{V}}{\partial\tau}\right\}dz \quad (7)$$
$$-\varepsilon^4\Delta_{\xi\eta}\text{rot}_{\xi\eta}\int_0^{2\pi}\vec{V}dz = 0$$

At the characteristic time scale $\tau \sim 1, t \sim \sqrt{R}$ we "freeze":

$$e^{-\varepsilon\tau} = b_0 \simeq 1$$

Now we can start seeking $\vec{V}_0$ and $\vec{V}_1$ in the form (6):

$$\varepsilon^0 : \text{from (4)} \longrightarrow \text{rot}_z[\text{rot}_z\vec{V}_0\times\vec{V}_0] + Ab_0\text{rot}_z[(\text{rot}_z\vec{V}_0-\vec{V}_0)\times\vec{h}_0] = 0$$

$$\text{div}_z\vec{V}_0 = 0$$

It can be easily seen, that

$$\vec{g}_0 = \vec{\delta}_0 = 0$$

Hence

$$\vec{V}_0(\xi,\eta,z,\tau) = \gamma_0^{(0)}(\xi,\eta,\tau)\vec{h}_0(z) + \gamma_0^{(1)}(\xi,\eta,\tau)\cdot\vec{h}_1(z)$$

Therefore

$$\text{div}_z\vec{V}_1 = -\text{div}_{\xi\eta}\vec{V}_0$$

$$\text{div}_z\vec{V}_1 = \frac{d}{dz}(B_1\cos z + C_1\sin z) = C_1\cos z - B_1\sin z$$

$$\text{div}_{\xi\eta}\vec{V}_0 = \left(\frac{\partial\gamma_0^{(0)}}{\partial\xi} - \frac{\partial\gamma_1^{(0)}}{\partial\eta}\right)\sin z + \left(\frac{\partial\gamma_0^{(0)}}{\partial\eta} + \frac{\partial\gamma_1^{(0)}}{\partial\xi}\right)\cos z,$$

$$\text{i.e. } B_1 = \left(-\frac{\partial\gamma_1^{(0)}}{\partial\eta} + \frac{\partial\gamma_0^{(0)}}{\partial\xi}\right) \quad ; \quad C_1 = -\left(\frac{\partial\gamma_0^{(0)}}{\partial\eta} + \frac{\partial\gamma_1^{(0)}}{\partial\xi}\right)$$

or

$$\vec{\vartheta}_1 = \begin{pmatrix} 0 \\ 0 \\ B_1 \cos z + C_1 \sin z \end{pmatrix} = \text{rot}_{\xi\eta}(\gamma_0^{(0)}\vec{h}_0(z) + \gamma_1^{(0)}\vec{h}_1(z)) \tag{8}$$

Since (8) holds obviously for $\vartheta_{n+1}, \gamma_0^{(n)}, \gamma_1^{(n)}, n \geq 1$, it enables the construction of the asymptotic expansion.

Set

$$\vec{V}_1 = \vec{w}_1 + \text{rot}_{\xi\eta}\vec{V}_0 + \vec{\delta}_1$$
$$\text{rot}_z \vec{V}_1 = \text{rot}_z \vec{w}_1 = \vec{w}_1$$
$$\vec{\delta}_1 = \begin{pmatrix} 0 \\ 0 \\ \delta_1(\xi,\eta,\tau) \end{pmatrix}$$

The substitution of $\vec{V}_0$ and $\vec{V}_1$ in this form into $\varepsilon^1$ and $\varepsilon^2$ terms of (4) yields

$$\frac{\partial \vec{V}_0}{\partial \tau} = \delta_1 \cdot \gamma_1^{(0)} \vec{h}_0 - \delta_1 \cdot (Ab_0 + \gamma_1^{(0)})\vec{h}_1$$

or

$$\begin{cases} \dfrac{\partial \gamma_0^{(0)}}{\partial \tau} = \delta_1 \cdot \gamma_1^{(0)} \\ \dfrac{\partial \gamma_1^{(0)}}{\partial \tau} = -\delta_1(Ab_0 + \gamma_0^{(0)}) \end{cases} \tag{9}$$

Therefore

$$\frac{1}{2}\frac{\partial}{\partial \tau}\left\{(\gamma_0^{(0)} + Ab_0)^2 + (\gamma_1^{(0)})^2\right\} = 0$$

or

$$(\gamma_0^{(0)} + Ab_0)^2 + (\gamma_1^{(0)})^2 = \text{const} = C_0^2(\xi,\eta) \tag{10}$$

Hence

$$\gamma_0^{(0)} + Ab_0 = C_0(\xi,\eta)\cos\varphi(\xi,\eta,\tau)$$
$$\gamma_1^{(0)} = C_0(\xi,\eta)\sin\varphi(\xi,\eta,\tau) \tag{11}$$

$$\delta_1 = -\frac{\partial \varphi}{\partial \tau} \tag{12}$$

Thus, according to (12), there might be an upward flow with velocity determined by the evolution of $\varphi$.

In order to get the evolutionary equation for $\delta_1$ (i.e., for $\varphi$), we have to substitute

$$\vec{V}_2 = \vec{w}_2 + \text{rot}_{\xi\eta}\vec{w}_1 + \vec{\delta}_2$$
$$\vec{V}_1 = \vec{w}_1 + \text{rot}_{\xi\eta}\vec{V}_0 + \vec{\delta}_1$$
$$\vec{V}_0 = \gamma_0^{(0)}\vec{h}_0 + \gamma_1^{(0)}\vec{h}_1$$

In to (7). It yields

$$\frac{\partial \vec{\delta}_1}{\partial \tau} = -\frac{1}{2}[\text{rot}_{\xi\eta}\text{rot}_{\xi\eta}\vec{V}_0 \times (\vec{V}_0 + Ab_0\vec{h}_0)]$$

However

$$\text{rot}_{\xi\eta}\text{rot}_{\xi\eta}\vec{V}_0 = -(\Delta_{\xi\eta}\gamma_0^{(0)})\vec{h}_0 - (\Delta_{\xi\eta}\gamma_1^{(0)})\vec{h}_1$$

and

$$[\vec{h}_0 \times \vec{h}_1] = \begin{pmatrix} 0 \\ 0 \\ -1 \end{pmatrix},$$

thus

$$\frac{\partial \delta_1}{\partial \tau} = \frac{1}{2}\left[(Ab_0 + \gamma_0^{(0)})\Delta_{\xi\eta}\gamma_1^{(0)}\right] - \frac{1}{2}\left[\gamma_1^{(0)}\Delta_{\xi\eta}\gamma_0^{(0)}\right] \qquad (13)$$

In view of (11), we get

$$\begin{cases} \dfrac{\partial^2 \varphi}{\partial \tau^2} = -\dfrac{C_0^2(\xi,\eta)}{2}\Delta_{\xi,\eta}\varphi + \left(\dfrac{\partial C_0}{\partial \xi}\dfrac{\partial \varphi}{\partial \xi} + \dfrac{\partial C_0}{\partial \eta}\dfrac{\partial \varphi}{\partial \eta}\right) \\ C_0^2(\xi,\eta) = (Ab_0 + \gamma_0^{(0)}(\xi,\eta,0))^2 + (\gamma_1^{(0)}(\xi,\eta,0))^2 \\ \varphi(\xi,\eta,0) = \text{arctg}\,\dfrac{\gamma_1^{(0)}(\xi,\eta,0)}{Ab_0 + \gamma_0^{(0)}(\xi,\eta,0)} \\ \dfrac{\partial \varphi}{\partial \tau}\bigg|_{\tau=0} = \delta_1(\xi,\eta,0) \equiv \delta_0(\xi,\eta) \end{cases} \qquad (14)$$

Thus, we have a Cauchy problem (with periodic boundary conditions) for an elliptic partial differential equation - the classic example of the "ill posed problem", i.e., **the case of instability in time**.

Since $\varepsilon^0$ and $\varepsilon^1$ approximations involve only the terms generated by the left-hand side of (1) or (2), we deal here with the **Eulerian instability**.

In order to elaborate $\vec{w}_1$ (the first term in $\vec{V}_1$) we have to substitute (6) for $n = 0, 1, 2$ into (4).

However, to do this we have to determine for which integer $\kappa$

$$\varepsilon^\kappa = R^{-1},$$

Suppose first, that $\kappa > 2$.

It yields for $\varepsilon^2$ approximation

$$\frac{\partial \vec{w}_1}{\partial \tau} + [\vec{w}_1 \times \vec{\delta}_1] + [(\vec{V}_0 + A\vec{h}_0) \times \vec{\delta}_2] = 0$$

To elucidate $w_1$, we similarly get the equations

$$\begin{cases} \dfrac{\partial \gamma_0^{(1)}}{\partial \tau} = \delta_1 \gamma_1^1 + \delta_2 \gamma_1^0 \\ \dfrac{\partial \gamma_1^{(1)}}{\partial \tau} = -\delta_1 \gamma_0^1 - \delta_2 (A + \gamma_0^0) \end{cases} \quad (15)$$

Consider the auxiliary system

$$\begin{cases} \dfrac{\partial \tilde{\gamma}_0^{(1)}}{\partial \tau} = \delta_1 \tilde{\gamma}_1^1 \\ \dfrac{\partial \tilde{\gamma}_1^{(1)}}{\partial \tau} = -\delta_1 \tilde{\gamma}_0^1 \end{cases} \quad (15')$$

Obviously

$$(\tilde{\gamma}_0^{(1)})^2 + (\tilde{\gamma}_1^{(1)})^2 = \tilde{C}_0^2(\xi, \eta),$$

Thus

$$\tilde{\gamma}_0^{(1)} = \tilde{C}_0 \cos \tilde{\varphi}(\xi, \eta, \tau)$$
$$\tilde{\gamma}_1^{(1)} = \tilde{C}_0 \sin \tilde{\varphi}(\xi, \eta, \tau),$$

Substituting this into (15'), we get

$$\delta_1 = -\frac{\partial \tilde{\varphi}}{\partial \tau}$$

Thus, in view of (12)

$$\tilde{\varphi} = \varphi(\xi, \eta, \tau) + \varphi_0(\xi, \eta)$$

Hence, we shall seek the solution of (15) as

$$\gamma_0^{(1)}(\xi,\eta,\tau) = \tilde{\tilde{C}}(\xi,\eta,\tau)\cos\tilde{\varphi}(\xi,\eta,\tau)$$

$$\gamma_1^{(1)}(\xi,\eta,\tau) = \tilde{\tilde{C}}(\xi,\eta,\tau)\sin\tilde{\varphi}(\xi,\eta,\tau)$$

Substituting this into (15), we get

$$\cos\tilde{\varphi}\frac{\partial\tilde{\tilde{C}}}{\partial\tau} = \delta_2\gamma_1^0$$

$$\sin\tilde{\varphi}\frac{\partial\tilde{\tilde{C}}}{\partial\tau} = -\delta_2(A+\gamma_0^0)$$

Multiplying the first equation by $A+\gamma_0^0 = C_0\cos\varphi$ and the second one by $\gamma_1^0 = C_0\sin\varphi$ we get

$$C_0\cos(\tilde{\varphi}-\varphi)\frac{\partial\tilde{\tilde{C}}}{\partial\tau} = 0$$

or

$$C_0\cos\varphi_0 \cdot \frac{\partial\tilde{\tilde{C}}}{\partial\tau} = 0$$

Then,

$$\delta_2 = 0$$

Hence

$$\tilde{C}^2(\xi,\eta) = (\gamma_1^{(1)}(\xi,\eta,0))^2 + (\gamma_0^{(1)}(\xi,\eta,0))^2$$

But the initial conditions for higher order terms in asymptotic expansion (5) are zero.

Thus

$$\tilde{C}_0(\xi,\eta) \equiv 0$$

and all the terms in the asymptotic expansion in form (6) for $n>1$ are vanishing. However, $\vec{V}_0 + \varepsilon(\vec{v}_1 + \vec{\delta}_1)$ is not an exact solution of (4).

Therefore, if we suppose that $\varepsilon = R^{-1/\kappa}$ for $\kappa > 2$, the solution of (4) in form (6) does not exist.

Set $\kappa = 2$, i.e. $\varepsilon = R^{-1/2}$, then the term $\frac{1}{R}\Delta_z(\text{rot}_z\vec{V})$ in (6) appears in $\varepsilon^2$ approximation and we get the following equation for $\vec{w}_1$:

$$\frac{\partial \vec{w}_1}{\partial \tau} + [(\vec{V}_0 + Ab_0\vec{h}_0) \times \vec{\delta}_2] + [\vec{w}_1 \times \vec{\delta}_1] + \vec{V}_0 = 0$$

To elucidate $\vec{w}_1$, we get the equations

$$\begin{cases} \dfrac{\partial \gamma_0^{(1)}}{\partial \tau} = \delta_1 \gamma_1^1 + \delta_2 \gamma_1^0 - \gamma_0^0 \\ \dfrac{\partial \gamma_1^{(1)}}{\partial \tau} = -\delta_1 \gamma_0^1 - \delta_2(A + \gamma_0^0) - \gamma_1^0 \end{cases} \quad (15'')$$

Consider the auxiliary system

$$\begin{cases} \dfrac{\partial \tilde{\gamma}_0^{(1)}}{\partial \tau} = \delta_1 \tilde{\gamma}_1^1 \\ \dfrac{\partial \tilde{\gamma}_1^{(1)}}{\partial \tau} = -\delta_1 \tilde{\gamma}_0^1 \end{cases} \quad (15''')$$

Obviously

$$(\tilde{\gamma}_0^{(1)})^2 + (\tilde{\gamma}_1^{(1)})^2 = \tilde{C}_0^2(\xi, \eta),$$

Thus

$$\tilde{\gamma}_0^{(1)} = \tilde{C}_0 \cos \tilde{\varphi}(\xi, \eta, \tau)$$
$$\tilde{\gamma}_1^{(1)} = \tilde{C}_0 \sin \tilde{\varphi}(\xi, \eta, \tau),$$

Substituting this into (15''), we get

$$\delta_1 = -\frac{\partial \tilde{\varphi}}{\partial \tau}$$

Thus, in view of (12)

$$\tilde{\varphi} = \varphi(\xi, \eta, \tau) + \varphi_0(\xi, \eta)$$

Hence, we shall seek the solution of (15) as

$$\gamma_0^{(1)}(\xi, \eta, \tau) = \tilde{\tilde{C}}(\xi, \eta, \tau) \cos \tilde{\varphi}(\xi, \eta, \tau)$$
$$\gamma_1^{(1)}(\xi, \eta, \tau) = \tilde{\tilde{C}}(\xi, \eta, \tau) \sin \tilde{\varphi}(\xi, \eta, \tau)$$

Substituting this into (15), we get

$$\cos \tilde{\varphi} \frac{\partial \tilde{\tilde{C}}}{\partial \tau} = \delta_2 \gamma_1^0 - \gamma_0^0$$

$$\sin \tilde{\varphi} \frac{\partial \tilde{\tilde{C}}}{\partial \tau} = -\gamma_1^0 - \delta_2(A + \gamma_0^0)$$

Multiplying the first equation by $A+\gamma_0^0 = C_0 \cos\varphi$ and the second one by $\gamma_1^0 = C_0 \sin\varphi$ we get

$$C_0 \cos(\tilde{\varphi}-\varphi)\frac{\partial \tilde{\tilde{C}}}{\partial \tau} = -[(\gamma_1^0)^2 + (A+\gamma_0^0)^2] + A(A+\gamma_0^0)$$

or

$$C_0 \cos\varphi_0 \cdot \frac{\partial \tilde{\tilde{C}}}{\partial \tau} = -C_0^2 + AC_0 \cos\varphi$$

However,

$$\varphi_0(\xi,\eta) = -\varphi(\xi,\eta,0) \neq 0$$

I.e. $\cos\varphi_0 > 0$ (small perturbation). Hence

$$\frac{\partial \tilde{\tilde{C}}}{\partial \tau} = -\frac{C_0}{\cos\varphi_0} + \frac{A\cos\varphi}{\cos\varphi_0} < 0,$$

i.e., $\tilde{\tilde{C}}(\xi,\eta,\tau)$ is a uniformly bounded decreasing function and $\vec{w}_1$ is negligible (in its impact on the streamlines formation) at a difference with $\delta_1$ that is growing exponentially with time.

Thus, we completed the asymptotic procedure for $\varepsilon^0$ and $\varepsilon^1$. Obviously, it can be done successively for any $n$, while equations (9) and (13) for $n>1$ will include as free terms $\gamma_k^{(0)}, \gamma_k^{(1)}, \delta_{k+1}$ for $k<n$, since these functions had been elucidated on the earlier stages and are considered as known.

Thus, in view of (8), we get

$$\vec{V} \sim \vec{V}_0 + \varepsilon \cdot (\vec{\delta}_1 + \vec{\vartheta}_1) = \gamma_0^0(\xi,\eta,\tau) \cdot \vec{h}_0(z) + \gamma_1^{(1)}(\xi,\eta,\tau)\vec{h}_1(z) +$$

$$+\varepsilon \cdot \begin{pmatrix} 0 \\ 0 \\ \delta_1(\xi,\eta,\tau) \end{pmatrix} + \varepsilon \cdot \begin{pmatrix} 0 \\ 0 \\ \left(-\frac{\partial \gamma_1^{(0)}}{\partial \eta} + \frac{\partial \gamma_0^{(0)}}{\partial \xi}\right)\cos z - \left(\frac{\partial \gamma_0^{(0)}}{\partial \eta} + \frac{\partial \gamma_1^{(0)}}{\partial \xi}\right)\sin z \end{pmatrix}$$

where:

$$\xi = x\cdot\varepsilon, \eta = y\cdot\varepsilon, \tau = t\cdot\varepsilon \qquad (16)$$

$$\varepsilon = R^{-1/2}$$

and

$$\vec{h}_0(z) = \begin{pmatrix} \sin z \\ \cos z \\ 0 \end{pmatrix} \quad ; \quad \vec{h}_1(z) = \begin{pmatrix} \cos z \\ -\sin z \\ 0 \end{pmatrix}$$

Hence,

$$\vec{u} = \vec{u}_0 + \tilde{V} \sim C_0 \cdot \begin{pmatrix} \sin(\varphi(\xi,\eta,\tau)+z) \\ \cos(\varphi(\xi,\eta,\tau)+z) \\ 0 \end{pmatrix} - R^{-1/2} C_0 \cdot \begin{pmatrix} 0 \\ 0 \\ \dfrac{\partial \varphi}{\partial \eta}\cos(\varphi+z) + \dfrac{\partial \varphi}{\partial \xi}\sin(\varphi+z) \end{pmatrix}$$

$$+ R^{-1/2} \cdot \begin{pmatrix} 0 \\ 0 \\ \dfrac{\partial C_0}{\partial \xi}\cos(z+\varphi) - \dfrac{\partial C_0}{\partial \eta}\sin(z+\varphi) - \dfrac{\partial \varphi}{\partial \tau} \end{pmatrix} \tag{17}$$

where

$$\xi = x \cdot R^{-1/2}, \quad \eta = y \cdot R^{-1/2}, \quad \tau = t \cdot R^{-1/2}$$

The first term in (17) might be called <u>a coherent Beltrami couple</u>, since it is a linear combination of two amplitude modulated in the $(x, y)$ plane Beltrami flows $\vec{h}_0(z)$ and $\vec{h}_1(z)$ of unit helicity, that exchange energy during the evolution, this exchange being controlled at all points $(x, y)$ by a <u>global phase function</u> $\varphi(\frac{x}{\sqrt{R}}, \frac{y}{\sqrt{R}}, \frac{t}{\sqrt{R}})$. Thus, for $\varphi = \frac{\pi}{2} + 2\pi\kappa$, the first term is in fact $C_0 \cdot \vec{h}_0(z)$, while for $\varphi = (2\kappa+1)\pi$, the first term becomes $C_0 \cdot \vec{h}_1(z)$.

While the first terms in (17) are of constant in time absolute value for $t \sim R^{1/2}$ and thus will decline at times $t > R$, due to the term $e^{-t/R}$ in equation (2), the last term will grow exponentially in $\tau \sim \frac{t}{\sqrt{R}}$ due to the fact that $\varphi$ is the solution of the equation (14). To elucidate the behavior of this term at very large times $t \gg R$, we consider the new scaling in time, leaving the space scaling (3) intact:

$$\frac{\partial}{\partial t} \to \varepsilon_1 \frac{\partial}{\partial \tau_1} \qquad (\tau_1 = \varepsilon_1 t, \varepsilon_1 \ll \frac{1}{R})$$

Looking for the expansion (5) in powers of $\varepsilon$, we follow the procedure as above. Since all the terms with $e^{\varepsilon\tau}$ will vanish at times $\tau \sim \tau_1 (e^{-\varepsilon\tau_1} = e^{-\varepsilon\varepsilon_1 t} \to 0)$ and $\varepsilon = R^{-1/2}$, we get

$$\text{rot}_z[\text{rot}_z\vec{V}\times\vec{V}] + \varepsilon\left\{\text{rot}_{\xi\eta}[\text{rot}_z\vec{V}\times\vec{V}] + \text{rot}_z[\text{rot}_{\xi\eta}\vec{V}\times\vec{V}]\right\} + \tag{4'}$$

$$+\varepsilon^2\left\{\text{rot}_{\xi\eta}[\text{rot}_{\xi\eta}\vec{V}\times\vec{V}]\Delta_z\text{rot}_z\vec{V}\right\} - \varepsilon^3\text{rot}_{\xi\eta}\Delta_z\vec{V} + \varepsilon^4\text{rot}_z\Delta_{\xi\eta}\vec{V} - \varepsilon^5\text{rot}_{\xi\eta}\Delta_{\xi\eta}\vec{V} +$$

$$+\varepsilon_1\frac{\partial}{\partial\tau_1}\left(\text{rot}_z\vec{V}\right) + \varepsilon\varepsilon_1\frac{\partial}{\partial\tau_1}\text{rot}_{\xi\eta}\vec{V} = 0$$

$$\text{div}_z\vec{V} + \varepsilon\,\text{div}_{\xi\eta}\vec{V} = 0$$

$$\varepsilon_1\frac{\partial}{\partial\tau_1}\int_0^{2\pi}\vec{V}dz + \varepsilon\int_0^{2\pi}[\text{rot}_{\xi\eta}\vec{V}\times\vec{V}]dz + \varepsilon^4\int\Delta_{\xi\eta}\vec{V}dz = 0 \tag{7'}$$

The asymptotic procedure yields

$$\vec{V} = \begin{pmatrix} 0 \\ 0 \\ V_0^{(3)}(\xi,\eta,\tau_1) \end{pmatrix}$$

Thus (dropping $\text{rot}_{\xi\eta}$), we get

$$\varepsilon\cdot\varepsilon_1\frac{\partial}{\partial\tau_{10}}\vec{V} = \varepsilon^5\Delta_{\xi\eta}\vec{V}_0$$

or

$$\varepsilon_1 = \varepsilon^4 = R^{-2}$$

Therefore,

$$\frac{\partial V_0^{(3)}}{\partial\tau_1} = \Delta_{\xi\eta}V_0^{(3)} \tag{14'}$$

The last term in (18) will serve as the initial condition for (14')

$$V_0^{(3)}(\xi,\eta,0) = \frac{\partial\varphi}{\partial\tau}\Big|_{\tau\sim 1}$$

with periodic boundary conditions, which is a parabolic equation whose bounded solutions decline exponentially in time $\tau_1 = \frac{t}{R^2}$.

Thus, the perturbed solutions of (2) would decline at times $t \sim R^2$.

## 2. Streamline Asymptotic Behaviour

Since, due to the first equations in (17) we have:

$$\frac{d\varphi(\xi(t),\eta(\tau),\tau)}{dt} = \frac{\partial \varphi}{\partial \xi}\cdot \dot{\xi}(\tau) + \frac{\partial \varphi}{\partial \eta}\cdot \dot{\eta}(\tau) + \frac{\partial \varphi}{\partial \tau} =$$

$$= C_0\left(\frac{\partial \varphi}{\partial \xi}\sin(\varphi + z) + \frac{\partial \varphi}{\partial \eta}\cos(\varphi + z)\right) + \frac{\partial \varphi}{\partial \tau},$$

the streamlines of the velocity field (17) are, in fact, the trajectories of the following system of ordinary differential equations:

$$\begin{cases} \dot{\xi} = C_0 \sin(\varphi(\xi,\eta,\tau) + z) \\ \dot{\eta} = C_0 \cos(\varphi(\xi,\eta,\tau) + z) \\ \dfrac{d(z+\varphi)}{d\tau} = \dfrac{\partial C_0}{\partial \xi}\cos(z+\varphi) - \dfrac{\partial C_0}{\partial \eta}\sin(z+\varphi) \end{cases} \quad (18)$$

$$\xi(0) = \xi_0, \eta(0) = \eta_0, z(0) = z_0$$

Here $\varphi(\xi,\eta,\tau)$ is defined by the Cauchy problem (14):

$$\begin{cases} \dfrac{\partial^2 \varphi}{\partial \tau^2} = -\dfrac{C_0^2(\xi,\eta)}{2}\left(\dfrac{\partial^2 \varphi}{\partial \xi^2} + \dfrac{\partial^2 \varphi}{\partial \eta^2}\right) + \dfrac{\partial C_0}{\partial \xi}\cdot \dfrac{\partial \varphi}{\partial \xi} + \dfrac{\partial C_0}{\partial \eta}\cdot \dfrac{\partial \varphi}{\partial \eta} \\ C_0^2(\xi,\eta) = (Ab_0 + \gamma_0^0(\xi,\eta))^2 + \gamma_1^0(\xi,\eta)^2 \\ \varphi(\xi,\eta,0) = \text{arctg}\dfrac{\gamma_1^0(\xi,\eta)}{Ab_0 + \gamma_0^0(\xi,\eta)} = \varphi_0(\xi,\eta) \\ \left.\dfrac{\partial \varphi}{\partial \tau}\right|_{\tau=0} = \delta_0(\xi,\eta) \end{cases}$$

where $\gamma_0(\xi,\eta)$ is the initial long-wavelength amplitude modulation of $\overrightarrow{h_0(z)}$ and $\gamma_1(\xi,\eta)$ is the initial long-wavelength amplitude modulation of $\overrightarrow{h_1(z)}$. The assumption of "long-wavelength perturbations" means that the functions $\gamma_0^{(0)}(\xi,\eta), \gamma_1^{(0)}(\xi,\eta), \delta_0(\xi,\eta)$ have finite number of terms in the two-dimensional Fourier expansion.

In original variables $x, y, z, t$ equations (19) for the streamlines of the secondary flow, generated by the long-wavelength instability of the non-stationary Trkal flow looks as:

$$\begin{cases} \dot{x} = C_0\left(\dfrac{x}{\sqrt{R}}, \dfrac{y}{\sqrt{R}}\right)\sin\left(z + \varphi\left(\dfrac{x}{\sqrt{R}}, \dfrac{y}{\sqrt{R}}, \dfrac{t}{\sqrt{R}}\right)\right) \\ \dot{y} = C_0\left(\dfrac{x}{\sqrt{R}}, \dfrac{y}{\sqrt{R}}\right)\cos\left(z + \varphi\left(\dfrac{x}{\sqrt{R}}, \dfrac{y}{\sqrt{R}}, \dfrac{t}{\sqrt{R}}\right)\right) \\ \dfrac{d(z+\varphi)}{dt} = \left(\dfrac{\partial C_0}{\partial x}\cos(z+\varphi) - \dfrac{\partial C_0}{\partial y}\sin(z+\varphi)\right)\cdot\dfrac{1}{\sqrt{R}} \end{cases} \quad (19)$$

where $C_0$ and $\varphi$ are defined by (14).

Set $w = z + \varphi$.

Consider the "quasi-stationary" solution $\bar{w}$ of the third equation in (18), defined as:

$$\frac{d\bar{w}}{d\tau} = 0$$

Thus

$$0 = \frac{d(\bar{z} + \varphi(\bar{\xi},\bar{\eta},\tau))}{d\tau} = \frac{\partial C_0(\bar{\xi},\bar{\eta})}{\partial \xi}\cos(\bar{z} + \varphi(\bar{\xi},\bar{\eta},\tau)) - \frac{\partial C_0(\bar{\xi},\bar{\eta})}{\partial \eta}\sin(\bar{z} + \varphi(\bar{\xi},\bar{\eta},\tau))$$

i.e.,

$$\frac{\dfrac{\partial C_0(\bar{\xi},\bar{\eta})}{\partial \xi}}{\dfrac{\partial C_0(\bar{\xi},\bar{\eta})}{\partial \eta}} = \text{tg}(\bar{z} + \varphi(\bar{\xi},\bar{\eta},\tau)) \quad (20)$$

Hence, due to the first and the second equations in (18):

$$\frac{d\bar{\xi}}{d\bar{\eta}} = \frac{(C_0(\bar{\xi},\bar{\eta}))_\xi}{(C_0(\bar{\xi},\bar{\eta}))_\eta} \quad (21)$$

Equation (21) determines some curve in $(\xi,\eta)$ plane. This curve is the projection of the "quasi-stationary" streamline $(\bar{\xi}(\tau), \bar{\eta}(\tau), \bar{z}(\tau))$ onto $(\xi,\eta)$ - plane,

$$\bar{z}(\tau) = -\varphi(\bar{\xi}(\tau), \bar{\eta}(\tau), \tau) + C_0$$

$$C_0 = z_0 + \varphi_0(\bar{\xi}_0, \bar{\eta}_0),$$

where $\varphi$ is a solution of (14), i.e., $\varphi$ is exponentially growing in time $\tau$.

It can be easily seen that $\bar{w}$ is a <u>stable solution</u> of the third equation in (18). In fact, set $w = \bar{w} + \tilde{w}$, where $\tilde{w}(\xi,\eta,z,\tau)$ is a small function,

$$\xi = \bar{\xi} + \tilde{\xi}, \eta = \bar{\eta} + \tilde{\eta}$$

Then, the linearization of the equations in (18) yields

$$\begin{cases} \dfrac{d\tilde{\xi}}{d\tau} = \left( \dfrac{\partial C_0}{\partial \xi} \tilde{\xi} + \dfrac{\partial C_0}{\partial \eta} \tilde{\eta} \right) \sin \bar{w} + \tilde{w} C_0(\bar{\xi}, \bar{\eta}) \cos \bar{w} \\[2mm] \dfrac{d\tilde{\eta}}{d\tau} = \left( \dfrac{\partial C_0}{\partial \xi} \tilde{\xi} + \dfrac{\partial C_0}{\partial \eta} \tilde{\eta} \right) \cos \bar{w} + \tilde{w} C_0(\bar{\xi}, \bar{\eta}) \sin \bar{w} \\[2mm] \dfrac{d\tilde{w}}{d\tau} = \dfrac{\partial}{\partial \xi}\left( \dfrac{\partial C_0}{\partial \xi} \tilde{\xi} + \dfrac{\partial C_0}{\partial \eta} \eta \right)\cos \bar{w} - \dfrac{\partial}{\partial \eta}\left( \dfrac{\partial C_0}{\partial \xi} \cdot \tilde{\xi} + \dfrac{\partial C_0}{\partial \eta} \tilde{\eta} \right)\sin \bar{w} - \tilde{w}\left( \dfrac{\partial C_0}{\partial \xi} \sin \bar{w} + \dfrac{\partial C_0}{\partial \eta} \cos \bar{w} \right) \end{cases}$$

However, we are interested in the perturbation vector $(\tilde{\xi}(\tau), \tilde{\eta}(\tau), \tilde{w}(\tau))$, which is orthogonal to the grad $C_0(\bar{\xi}, \bar{\eta})$ vector, i.e.

$$\left( \dfrac{\partial C_0}{\partial \xi} \right)_{\substack{\xi=\bar{\xi}\\\eta=\bar{\eta}}} \cdot \tilde{\xi} + \left( \dfrac{\partial C_0}{\partial \eta} \right)_{\substack{\xi=\bar{\xi}\\\eta=\bar{\eta}}} \cdot \tilde{\eta} = 0$$

Thus,

$$\begin{cases} \dfrac{d\tilde{\xi}}{d\tau} = \tilde{w} C_0(\bar{\xi}, \bar{\eta}) \cos \bar{w} \\[2mm] \dfrac{d\tilde{\eta}}{d\tau} = -\tilde{w} C_0(\bar{\xi}, \bar{\eta}) \sin \bar{w} \\[2mm] \dfrac{dw}{d\tau} = -\left[ \dfrac{\partial C_0}{\partial \xi} \sin \bar{w} + \dfrac{\partial C_0}{\partial \eta} \cos \bar{w} \right] \tilde{w} \end{cases}$$

where $\operatorname{tg}\bar{w} = \dfrac{\partial C_0}{\partial \xi} \Big/ \dfrac{\partial C_0}{\partial \eta}$.

However, this means that

$$\sin \bar{w} = \dfrac{\partial C_0}{\partial \xi} \cdot \dfrac{1}{\sqrt{(\frac{\partial C_0}{\partial \xi})^2 + (\frac{\partial C_0}{\partial \eta})^2}}$$

$$\cos \bar{w} = \dfrac{\partial C_0}{\partial \eta} \cdot \dfrac{1}{\sqrt{(\frac{\partial C_0}{\partial \xi})^2 + (\frac{\partial C_0}{\partial \eta})^2}},$$

hence,

$$\dfrac{\partial C_0}{\partial \xi} \sin \bar{w} + \dfrac{\partial C_0}{\partial \eta} \cos \bar{w} = \sqrt{(\dfrac{\partial C_0}{\partial \xi})^2 + (\dfrac{\partial C_0}{\partial \xi})^2}$$

and

$$\frac{d\tilde{w}}{d\tau} = -\sqrt{(\frac{\partial C_0}{\partial \xi}) + (\frac{\partial C_0}{\partial \eta})^2} \cdot \tilde{w}$$

Since in the bounded domain we have (besides small vicinities of a finite number of stationary points, where $\text{grad} C = 0$)

$$m \leq \sqrt{(\frac{\partial C_0}{\partial \xi})^2 + (\frac{\partial C_0}{\partial \xi})^2} \leq M,$$

we have

$$|\tilde{w}_{(t)}| = c_0 e^{-\int_{\tau_0}^{\tau} \sqrt{(\frac{\partial C_0}{\partial \xi})^2 + (\frac{\partial C_0}{\partial \xi})^2} d\tilde{\tau}} \leq C_0 e^{-m(\tau - \tau_0)}$$

Hence, the projection of the streamline onto the $(\xi, \eta)$ - plane is a curve that approaches asymptotically to some "limit curve" defined by the equations for quasi-stationary solutions:

$$\dot{\bar{\xi}} = C_0 \cdot \frac{\partial C_0}{\partial \xi} \cdot \frac{1}{|\text{grad} C_0(\bar{\xi}, \bar{\eta})|}$$
$$\dot{\bar{\eta}} = C_0 \cdot \frac{\partial C_0}{\partial \eta} \cdot \frac{1}{|\text{grad} C_0(\bar{\xi}, \bar{\eta})|}$$
(21′)

Or, in the vector form

$$\dot{\overline{\Sigma}} = C_0(\overline{\Sigma}) \cdot \frac{\text{grad} C_0(\overline{\Sigma})}{|\text{grad} C_0(\overline{\Sigma})|} \quad (22)$$

where

$$\overline{\Sigma} = \begin{pmatrix} \bar{\xi} \\ \bar{\eta} \end{pmatrix}$$

Thus, the projection of the "quasi-stationary" streamline onto the $(\xi, \eta)$ plane "attracts" the streamlines, defined by (18) as $\tau \to \infty$, i.e. these streamlines are "attracted" by some surface in $(\xi, \eta, z)$ - space, defined by (21).

Hence, "large scale structures" emerge as the integral lines of the gradient vector field for the function

$$C_0(\xi, \eta) = \left[ (Ab_0 + \gamma_0(\xi, \eta))^2 + \gamma_1^2(\xi, \eta) \right]^{1/2}$$

where $\gamma_0(\xi,\eta)$ and $\gamma_1(\xi,\eta)$ are initial small long-wave amplitude modulations ("noise") of the Beltrami flows

$$\vec{h}_0(z) = \begin{pmatrix} \sin z \\ \cos z \\ 0 \end{pmatrix} \quad ; \quad \vec{h}_1(z) = \begin{pmatrix} \cos z \\ -\sin z \\ 0 \end{pmatrix}$$

of unit helicity and the projection of the "quasi-stationary" streamline onto the $(\xi,\eta)$ plane is the integral line of the grad $C_0(\xi,\eta)$ vector field.

Consider the evolution of the $C_0(\bar{\xi}(\tau),\bar{\eta}(\tau)) = C_0(\tau)$ along the quasi-stationary streamline $(\bar{\xi}(\tau),\bar{\eta}(\tau))$:

$$\frac{dC_0(\tau)}{d\tau} = \frac{\partial C}{\partial \xi}\cdot\dot{\bar{\xi}}(\tau) + \frac{\partial C}{\partial \eta}\dot{\bar{\eta}}(\tau) = \frac{\partial C}{\partial \xi}\cdot C_0(\bar{\xi},\bar{\eta})\sin\bar{w} + \frac{\partial C_0}{\partial \eta}\cdot C_0(\bar{\xi},\bar{\eta})\cos\bar{w} =$$
$$= C_0(\bar{\xi},\bar{\eta})\cdot\sqrt{(\frac{\partial C_0}{\partial \xi})^2 + (\frac{\partial C_0}{\partial \eta})^2}$$

or

$$\frac{dC_0(\tau)/d\tau}{C_0(\tau)} = \sqrt{(\frac{\partial C_0}{\partial \xi})^2 + (\frac{\partial C_0}{\partial \eta})^2}$$

Thus,

$$|C_0(\tau)| = c_0 \exp\int_0^\tau \sqrt{(\frac{\partial C_0(\xi(t),\eta(t))}{\partial \xi})^2 + (\frac{\partial C_0(\xi(t),\eta(t))}{\partial \eta})^2}\,dt \qquad (3)$$

Since $C_0(\xi,\eta)$ is a bounded function in the whole plane, the quasi-stationary trajectory can remain beyond any given vicinity of the stationary points (22) only for a limited time span. Thus, we proved the well known fact, that the integral lines of the gradient field connect the "stationary points", where the gradient $C_0(\xi,\eta)$ vanishes:

$$|\text{grad}C_0(\bar{\xi},\bar{\eta})| = 0.$$

Thus, "large scale structures" are formed from these stable $(t \sim \sqrt{R})$ curves in the $(x, y)$ plane. Hence, the question of the streamline behaviour under long wavelength perturbations of the Trkal solution for the forceless Navier-Stokes equation is reduced to the determination of the gradient lines for the function

$$C_0 = [(A + \gamma_0(\xi,\eta))^2 + \gamma_1^2(\xi,\eta)]^{1/2},$$

where $\gamma_0(\xi,\eta)$ and $\gamma_1(\xi,\eta)$ are the initial large scale amplitude modulations of the Beltrami flows $\overrightarrow{h_0(z)}$ and $\overrightarrow{h_1(z)}$, both being of unit vorticity, while

$$\xi \sim \frac{x}{\sqrt{R}} \quad \text{and} \quad \eta \sim \frac{y}{\sqrt{R}}.$$

The stationary points of $C_0(\xi,\eta)$ are either points of maximum ("sources") or minimum values ("syncs") or saddle points.

Every trajectory starts at some maximal point and ends at some minimal point. The saddle point has one incoming and one outgoing trajectory - the separatrices. The plane domain is thus partitioned by the separatrices into sub domains containing the trajectories (plane streamlines), which connect one maximum critical point with one minimum critical point, while there are no other critical points inside these sub domains, i.e. the trajectories inside the sub domains are homotopic.

It can be easily seen from (6),(8) and (17), that

$$\text{u-rotu} = (1/R^{1/2})\vec{\delta}_1 + O(1/R),$$

where the last term in the right hand side contains all the terms of the higher orders in the asymptotic expansion, while the first term in the right hand-side is a vector, which is parallel to the z –axis. Thus, up to terms of order 1/R, both velocity and vorticity vectors belong to a tangent plane of a vertical surface, which contains a curve in the (x, y) plane, determined by the equations (21'). This is true even, when the first term in the right-hand side of the last equation is not small, i.e. when velocity and vorticity cease to be almost collinear. Thus, this surface might be considered as "velocity sheet", as well as "vortex sheet". Velocity sheets, which contain homotopic quasi stationary trajectories of the same sub domain in (x, y) plane, connecting two fixed critical points, form the invariant three dimensional domains, called "velocity tubes". Clearly, up to terms of order 1/R, "velocity tubes" are at the same time "vortex tubes". These velocity –vorticity tubes will be stable at times $\tau \sim 1, t \sim \sqrt{R}$.

## 3. Qualitative (topological) Example

For the sake of demonstration of the gradient lines concept we consider here one simple case.

The assumption of the existence of the "long-wave perturbations" means that $\gamma_0(\xi,\eta)$ and $\gamma_1(\xi,\eta)$ are two-periodic functions in $(\xi,\eta)$. Otherwise all the space frequencies would be present in the Fourier expansion of these functions and the perturbation cannot be called "long-wave". Hence, $\gamma_0(\xi,\eta)$ and $\gamma_1(\xi,\eta)$ are the trigonometric polynomials of two variables with finite number of terms.

<u>We shall limit ourselves to the case when only the first harmonics are present</u>. Since $\gamma_0(\xi,\eta)$ and $\gamma_1(\xi,\eta)$ are small fluctuations we may introduce the parameter of smallness $\bar{\varepsilon}$, which actually might be the biggest of the maximums of the absolute values. Then

$$C_0 = [(A + \bar{\varepsilon}\bar{\gamma}_0(\xi,\eta))^2 + \bar{\varepsilon}^2\bar{\gamma}_1^2(\xi,\eta)]^{1/2}$$
$$= [(A^2 + 2\bar{\varepsilon}\bar{\gamma}_0(\xi,\eta))]^{1/2}(1 + \frac{\varepsilon^2}{2A}(\bar{\gamma}_0^2 + \bar{\gamma}_1^2) + \cdots)$$

It can be proved by means of algebraic topology [25], that for sufficiently small $\bar{\varepsilon}$, the picture of the gradient lines for the function $C_0$ is topologically equivalent to the picture of the gradient lines for the function

$$\tilde{C}_0^2 = A^2 + 2\bar{\varepsilon}\bar{\gamma}_0(\xi,\eta)$$

The simplest case concerns $2\pi$ periodic conditions on the square $[0,2\pi]\times[0,2\pi]$ for $\gamma_0(\xi,\eta)$ (i.e. $\gamma_0(\xi,\eta)$ is considered on the torus), having first harmonics exclusively in the Fourier expansion. Arnold [1] writes these functions in the form

$$\bar{\gamma}_0(\xi,\eta) = a\cos\xi + b\sin\xi + c\cos\eta + d\sin\eta +$$
$$+ p\cos(\xi+\eta) + q\sin(\xi+\eta)$$

and proves that they have six stationary points and they allow two different topological pictures (with respect to the diffeomorphism group of the torus) for the level lines. These two pictures (and consequently the gradient line pictures) are determined by the structure of the six stationary points:

- two maximal points, three saddle points and one minimal point;

- one maximal point, three saddle points and two minimal points

The corresponding picture of the gradient lines for Case b) is presented in Figure 1.

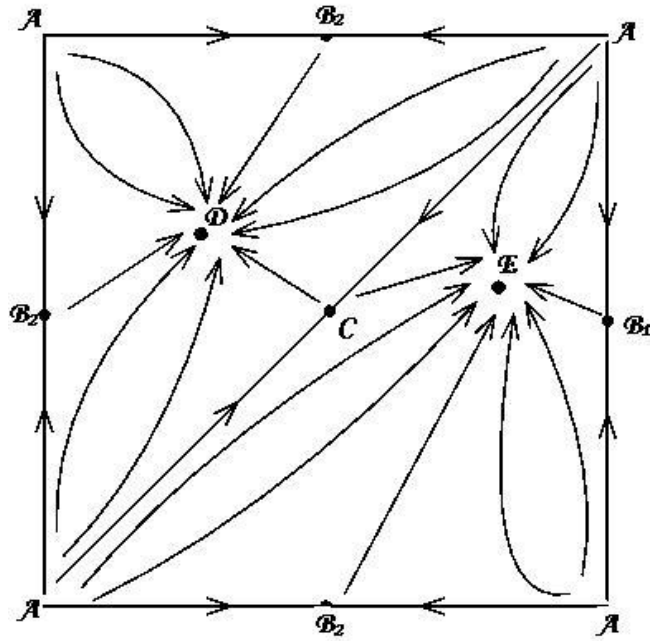

Figure 1

The related three dimensional picture to the above studied case is given on Fig.2

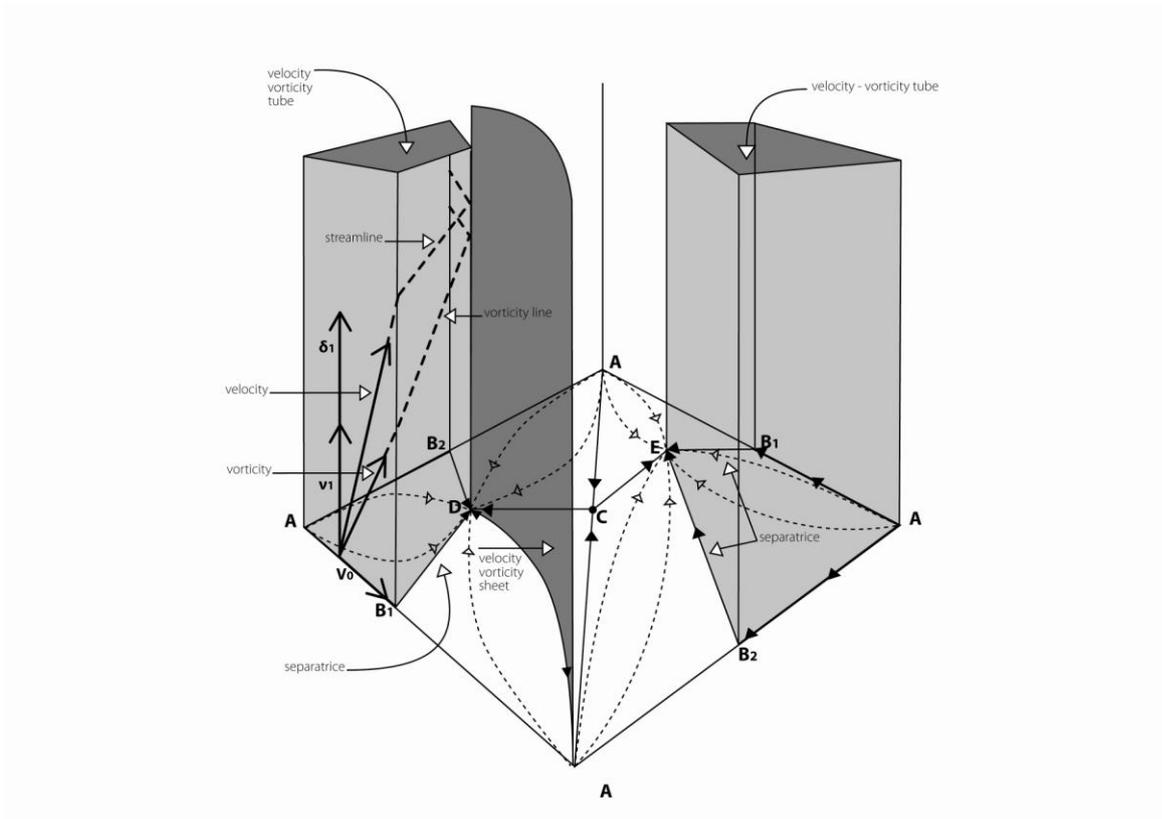

**Fig.2**

## Conclusions.

Multi scaling asymptotic analysis at large values of the Reynolds number R is applied to the long wavelength perturbations of the non stationary anisotropic helical solution for the forceless Navier Stokes equation. This allows to discern specific patterns in the orthogonal plane to the anisotropy direction. These patterns appear as the gradient lines of some function determined by the initial conditions for large scale perturbations of two Beltrami flows of the same helicity.

The eventual coupling of these amplitude modulated Beltrami flows yields the emergence of quasi stationary patterns at times of order $R^{1/2}$. These patterns might be interpreted as large scale structures of the characteristic size $R^{1/2}$. The energy density of the flow will be eventually a function that varies on the large characteristic space scale of order $R^{1/2}$. Such phenomenon was envisaged in [5] in relation to geophysical phenomena. The fact of formation of large scale Beltrami like helical structures, tornadoes, tropical storms, cloud streets, etc. is firmly established by observations [27]

The general picture is determined by the basic features of the initial conditions and sometimes it might be elucidated without complex computations. Similar (homotopic) plane flow patterns are bounded by the separatrices, which determine the invariant sub domains in the plane as well as stable invariant three dimensional velocity tubes, which are at the same time (up to terms of order 1/R) the vortex tubes, i.e. they might be considered as consisting of velocity lines as well as vortex lines .

The 3-dimensional large scale flow velocity has a non stable component in the anisotropic direction, which can be directly discerned as the solution of the Cauchy problem ("ill posed problem") for an elliptic partial differential equation with coefficients determined by the initial conditions. The upward flow outlives the basic Trkal flow and vanishes at times of order $R^2$.

## Acknowledgements

The author is deeply grateful to Professor E. Levich for numerous introspects into the modern concepts of turbulence, and for the relentless attention to the author's efforts in the preparation of the present work. The author is also deeply grateful to Professor L. Polterovich for consultations. Finally thanks to Ms. Bette Lewis for multiple editing and typing of this manuscript.